\documentclass[preprint2]{exoplanet}
\usepackage{times}

\newcommand{\mj}{$M_{\mathrm{J}}$}
\newcommand{\rj}{$R_{\mathrm{J}}$}
\newcommand{\me}{$M_{\oplus}$}

\newcommand{\hd}{HD 209458b} 
\newcommand{\hh}{HD 149026b}

\newcommand{\teff}{$T_{\rm eff}$}

\newcommand{\cp}{\citep}
\newcommand{\ct}{\citet}

\def\msol{M_\odot}

\def\mjup{M_{\rm J}}
\def\rjup{R_{\rm J}} 

\def\mearth{\,{\rm M}_\oplus}

\def\mcore{\,{\rm M}_{\rm core}}

\def\te{T_{\rm eff}}

\def\simgr{\,\hbox{\hbox{$ > $}\kern -0.8em \lower 1.0ex\hbox{$\sim$}}\,}
\def\simle{\,\hbox{\hbox{$ < $}\kern -0.8em \lower 1.0ex\hbox{$\sim$}}\,}
\def\aap{{\em Astron. Astrophys. }}
\def\apj{{\em Astrophys. J.}}
\def\beq{\begin{equation}}
\def\eeq{\end{equation}}

\begin{document}

\title{\textbf{\LARGE Giant Planet Interior Structure and Thermal Evolution}}

\author {\textbf{\large Jonathan J. Fortney}}
\affil{\small\em University of California, Santa Cruz}
\author {\textbf{\large Isabelle Baraffe}}
\affil{\small\em Ecole Normale Sup\'erieure de Lyon - CRAL}
\author {\textbf{\large Burkhard Militzer}}
\affil{\small\em University of California, Berkeley}

\begin{abstract}
\begin{list}{ } {\rightmargin 0.5in}{\leftmargin 1in}
\baselineskip = 11pt
\parindent=1pc
{\small We discuss the interior structure and composition of giant planets, and how this structure changes as these planets cool and contract over time.  Here we define giant planets as those that have an observable hydrogen-helium envelope, which includes Jupiter-like planets, which are predominantly H/He gas, and Neptune-like planets which are predominantly composed of elements heavier than H/He.  We describe the equations of state of planetary materials and the construction of static structural models and thermal evolution models.  We apply these models to transiting planets close to their parent stars, as well as directly imaged planets far from their parent stars.  Mechanisms that have been postulated to inflate the radii of close-in transiting planets are discussed.  We also review knowledge gained from the study of the solar system's giant planets.  The frontiers of giant planet physics are discussed with an eye towards future planetary discoveries.
 \\~\\~\\~}
 
\end{list}
\end{abstract}

\section{INTRODUCTION}

The vast majority of planetary mass in the solar system, and indeed the galaxy, is hidden from view in the interiors of giant planets.  Beyond the simple accounting of mass, there are many reasons to understand these objects, which cut across several disciplines.  Understanding the structure of these planets gives us our best evidence as to the formation mode of giant planets, which tells us much about the planet formation process in general.  As we will see, giant planets are vast natural laboratories for simple materials under high pressure in regimes that are not yet accessible to experiment.  With the recent rise in number and stunning diversity of giant planets, it is important to understand them as a class of astronomical objects.

We would like to understand basic questions about the structure and composition of giant planets.  Are they similar in composition to stars, predominantly hydrogen and helium with only a small mass fraction ($\sim$1\%) of atoms more massive than helium?  If giant planets are enhanced in  ``heavy elements'' relative to stars, are the heavy elements predominantly mixed into the hydrogen-helium (H-He) envelope, or are they mainly found in a central core?  If a dense central core exists, how massive is it, what is its state (solid or liquid), and is it distinct or diluted into the above H-He envelope?  Can we understand if a planet's heavy element mass fraction depends on that of its parent star?

Giant planets are natural laboratories of hydrogen and helium in the megabar to gigabar pressure range, at temperatures on the order of 10$^4$ K. How do hydrogen and helium interact under these extreme conditions? Is the helium distribution within a planet uniform and what does this tell us about how H and He mix at high pressure?  What methods of energy transport are at work in the interiors of giant planets?  Can we explain planets' observable properties such as the luminosity and radius at a given age?

The data that we use to shape our understanding of giant planets comes from a variety of sources.  Laboratory data on the equation of state (EOS, the pressure-density-temperature relation) of hydrogen, helium, warm fluid ``ices'' such at water, ammonia, and methane, silicate rocks, and iron serve at the initial inputs into models.  Importantly, data are only avaible over a small range of phase space, so that detailed theoretical EOS calculations are critical to understanding the behavior of planetary materials at high pressure and temperature.  Within the solar system, spacecraft data on planetary gravitational fields allows us to place constraints on the interior density distribution for Jupiter, Saturn, Uranus, and Neptune.  For exoplanets, we often must make due with far simpler information, namely a planet's mass and radius only.  For these distant planets, what we lack in detailed knowledge about particular planets, we can make up for in number.

Within six years of the Voyager 2 fly-by of Neptune, the encounter that completed our detailed census of the outer solar system, came the stunning discoveries of the extrasolar giant planet 51 Peg b \cp{Mayor95} and also the first bona fide brown dwarf, Gliese 229B \cp{Nakajima95}.  We were not yet able to fully understand the structure and evolution of the solar system's planets before we were given a vast array of new planets to understand.  In particular the close-in orbit of 51 Peg b led to immediate questions regarding its history, structure, and fate \cp{Guillot96,Lin96}.  Four years later, the first transiting planet, \hd\ \cp{Charb00,Henry00}, was found to have an inflated radius of $\sim$1.3 \rj, confirming that proximity to a parent star can have dramatic effects on planetary evolution \cp{Guillot96}.  The detections of over 50 additional transiting planets (as of August 2009) has conclusively shown that planets with masses greater than that of Saturn are composed predominantly of H/He, as expected.  However, a great number of important questions have been raised.

Much further from their parent stars, young luminous gas giant planets are being directly imaged from the ground and from space \cp{Kalas08,Marois08,Lagrange09}.  For imaged planets, planetary thermal emission is only detected in a few bands, and a planet's mass determination rests entirely on comparisons with thermal evolution models, that aim to predict a planet's luminosity and spectrum with time.  However, the luminosity of young planets is not yet confidently understood \cp{Marley07,Chabrier07}.

In this chapter we first outline the fundamental physics and equations for understanding the structure and thermal evolution of giant planets.  We next describe the current state of knowledge of the solar system's giant planets.  We then discuss current important issues in modeling exoplanets, and how models compare to observations of transiting planets, as well as directly imaged planets.  We close with a look at the future science of extrasolar giant planets (EGPs).

\bigskip
\section{EQUATIONS AND MODEL BUILDING}
\bigskip

\noindent
\textbf{ 2.1 Properties Of Materials At High Pressure} \label{phys}
\bigskip

Materials in the interiors of giant planets are exposed to extreme
temperature and pressure conditions reaching $\sim$10000 K and
100--1000 GPa (1-10 Mbar). (See Figure 1.) The characterization of materials' properties under
these conditions has been one of the great challenges in experimental
and theoretical high pressure physics. Ideally one would recreate such
high pressure in the laboratory, characterize the state of matter, and
then directly measure the equation of state (EOS) as well as the
transport properties needed to model planetary interiors. While a
number of key experiments have been performed, for a large part of
Jupiter's interior we instead rely on theoretical methods.

There are both static and dynamic methods to reach high pressures in
laboratory experiments. The highest pressure in static compression
experiments have been reached with diamond anvil
cells~\cp{Hemley1998,GeneralReview,Review_H}. One has been able to
reach $\sim$400 GPa (4 Mbar), which exceeds the pressure in the center of the
Earth but far less from $\sim$4000 GPa at Jupiter's core-envelope 
boundary. While many diamond anvil cell experiments were performed at
room temperature, the combination with laser heating techniques has
enabled one to approach some of temperature conditions that exist
inside planets.

\begin{figure*}
\epsscale{1.2}
\plotone{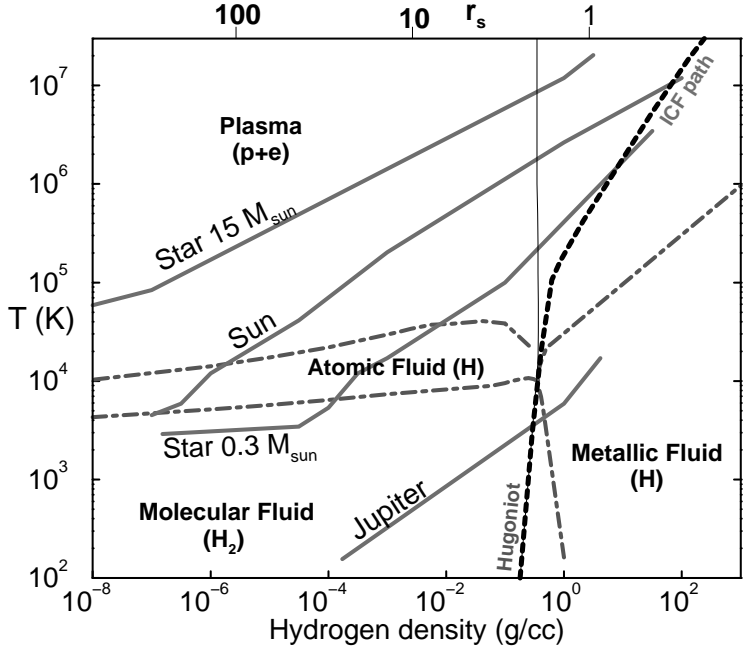}
\caption{Density-temperature phase diagram of hot dense hydrogen. The dash-dotted lines separate the molecular, atomic, metallic, and plasma regimes. The solid lines are isentropes for Jupiter and stars with 0.3, 1, and 15 solar masses.  Single shock Hugoniot states as well as the inertial confinement fusion paths are indicated by dashed lines. The thin solid line shows \emph{$\rho$-T} conditions of PIMC simulations.
\label{Prho}}	
\end{figure*}

The challenges in dynamic compression experiments are quite
different. Reaching the required pressures is not the primary
concern, but instead it is the difficulty in achieving a high enough
density. Most dynamic experiments compress the material with a single
shock wave. The locus of final $\rho-T-P$ points reached by single shock from one particular initial point is called the Hugoniot.  While this method provides direct access to the
EOS~\cp{Ze66}, the compression ratio rarely exceed values of
4. Instead the material is heated to very high temperature that exceed
those in planetary interiors~\cp{Jeanloz07,MH08}. Recently, static and
dynamic compression techniques have been combined to address this
issue~\cp{eggert08}. By precompressing the sample in a diamond anvil
cell before a shock wave was launched, Eggert et al.~were able to probe
deeper into planetary interiors. Earlier experiments by~\ct{We96}
employed reverberating shock waves to reach high densities and
thereby approached the state of metallic hydrogen at high temperature.

The first laser shock experiments that reached megabar pressures
predicted that the material to be highly compressible under shock
conditions and to reach densities six times higher than the initial
state~\cp{Si97,Co98}. However, later
experiments~\cp{Knudson01,Kn03,Belov2002,BoriskovNellis05} showed smaller
compression ratios of about 4.3, which were in good agreement with
theoretical predictions~\cp{Le97,MC00}.

Until very recently all models for giant planet interiors were based
on {\em chemical models}~\cp{ER85a,SC92,SC95,Ju00} that describe
materials as a ensemble of stable molecules, atoms, ions, and free
electrons. Approximations are made to characterize their
interactions. The free energy of the material is calculated with
semi-analytical techniques and all other thermodynamic variables are
derived from it. Chemical models require very little computer time, can
easily cover orders of magnitude in pressure-temperature space, and
have therefore been applied to numerous star and planet
models. Chemical models do not attempt to characterize all the
interactions in a many-body system. Molecular hydrogen at high
density, e.g., is typically approximated by a system of hard spheres
where the excitation spectrum of the isolated molecule is modified by a
density dependent term.  In general, chemical models have difficulties
predicting materials' properties in the strongly coupled regime where
interaction effects dominated over kinetic effects. At high density and
temperature where molecules are no longer stable or near a
metal-to-insulator transition, chemical models require input from
experiments or other theoretical techniques to fit adjustable
parameters.

Recently, progress in the field of theoretical description of
planetary materials at high pressure has come from {\em first
principles} computer simulations. Such methods are based on the
fundamental properties of electrons and nuclei and do not contain any
parameters that are fit to experimental data. While approximations
cannot be avoided altogether to efficiently derive a solution to the
many-body Schr{\"o}dinger equation, such approximations are not
specific to a particular material and have been tested for a wide
range of materials and different thermodynamic conditions. 

First-principles simulation can now routinely study the behavior of
hundreds of particles at very different pressure and temperature
conditions. Here we summarize three different approaches: path
integral Monte Carlo (PIMC), density functional molecular dynamics
(DFT-MD), as well as quantum Monte Carlo (QMC). The challenge of
performing accurate simulations has always been to make sure that the
approximations, that are often necessary to perform the calculations
at all, do not impact the predictions in a significant way. There are
{\em fundamental approximations} such as the assumption of simplified
functionals in DFT or the nodal approximation in QMC and PIMC
simulations with fermions~\cp{FM99}. These approximations can in most
cases only be checked by comparison between different methods or with
experimental results. Then there are also {\em controlled
approximations} such as using a sufficiently large number of particles,
long enough simulations, or a large enough basis set. These
approximations can always be verified by investing additional computer
time but it is not always possible to perform all tests at all
thermodynamic conditions.

\begin{figure}[!]
\epsscale{.80}
\plotone{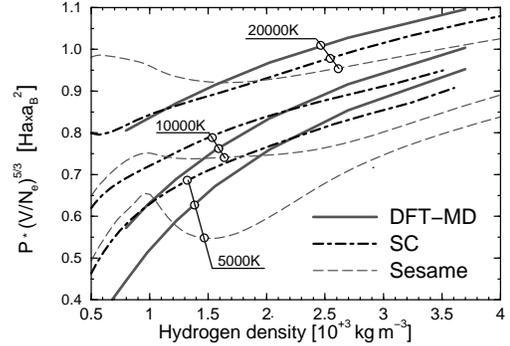}
\caption{Comparison of the DFT-MD EOS with the Saumon-Chabrier (SC) and Sesame models. Three isotherms
       for pure hydrogen are shown in the metallic regime at high
       pressure. $P$ is scaled by the volume per electron to the power
       $\frac{5}{3}$ to remove most of the density dependence.
\label{P_vs_rho}}	
\end{figure}

PIMC~\cp{PC84,Ce95} is a finite-temperature quantum simulation method
that explicitly constructs paths for electrons and nuclei. All
correlation effects are included, which makes PIMC one of the most
accurate finite-temperature quantum simulation methods available. The
only fundamental approximation required is the fixed node
approximation~\cp{Ce91,Ce96} that is introduced to treat the fermion
sign problem, which arises from the explicit treatment of
electrons. The method is very efficient at high temperature and can
provide one coherent description of matter reaching up to a fully
ionized plasma state. PIMC simulations have been applied to
hydrogen~\cp{PC94,MC00,MC01}, helium~\cp{Mi06,Mi08c}, and their
mixtures~\cp{Mi05}. At low temperature, this method becomes more
computationally demanding because the length of the path scales like
1/$T$. At temperatures below $\sim$5000$\,$K where electronic
excitations are not important, it is more efficient to use a
ground-state simulation method discussed next.

Density functional molecular dynamics simulations rely on the
Born-Oppenheimer approximation to separate the motion of electrons and
nuclei. For a given configuration of nuclei, the instantaneous
electron ground-state is derived from density functional theory. Forces
are derived and nuclei are propagated using classical molecular
dynamics. Excited electronic states can incorporated by using the Mermin
functional~\cp{Mermin65}. 

DFT is a mean field approach and approximations are made to treat
electronic exchange and correlation effects. Electronic excitations
gaps are underestimated in many materials. A more sophisticated
description of electronic correlation effects is provided by quantum
Monte Carlo~\cp{FM99} where one uses an ensemble of random walks to
project out the ground-state wave function. This method represents the
ground-state analog of PIMC. To avoid the fermion sign problem, one
also introduces a nodal approximation based on a trial wave function.
While most QMC calculations were performed for fixed nuclei, the
method has recently been extended to fluids and calculations for fluid
hydrogen have been performed~\cp{PCH04,delaney06}.

The EOS of dense hydrogen has been the subject of several DFT-MD
studies~\cp{Le97,Desjarlais2003,Bonev2004}. Fig.~\ref{P_vs_rho}
compares DFT-MD EOS from~\ct{MHVTB} with predictions from free
energy models. Even at the highest densities, one finds significant
deviations because no experimental data exist to guide free energy
models. Furthermore, the density is not yet high enough for hydrogen to
behave like an ideal Fermi gas.

\begin{figure}[!]
\epsscale{.80}
\plotone{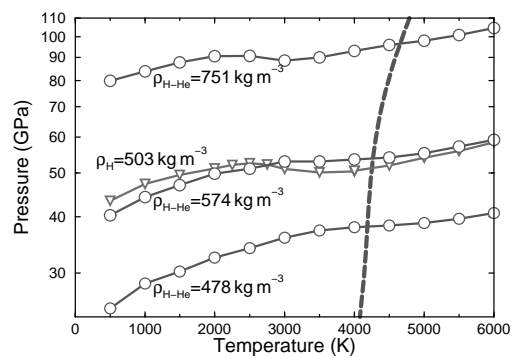}
\caption{Isochores derived from DFT-MD simulations of
       H-He mixtures (circles, $Y$=0.2466) and pure hydrogen
       (triangles). Results for mixtures predict a positive
       Gr\"uneisen parameter, $(\partial P/\partial T)|_V > 0$, along
       Jupiter's isentrope.
\label{P_vs_T}}	
\end{figure}

Of importance for the interiors of giant planets and the generation of
their magnetic fields are the properties of the insulator-to-metal
transition in dense hydrogen~\cp{Chabrier2006}. According to
predictions from the best simulation methods currently available,
quantum Monte Carlo~\cp{delaney06} and DFT-MD~\cp{Vo07}, this
transition is expected to occur \emph{gradually} in the condition of giant
planet interiors. Earlier DFT-MD
simulations~\cp{Scandolo2003,Bonev2004} had predicted a sharp
dissociation transition but all these results have now been attributed
to inaccuracies in the wave function propagation with the
Car-Parinello method~\cp{CP85}. With the more accurate
Born-Oppenheimer propagation method, the transition occurs gradually
(see the discussion in~\ct{hedla2008}) but gives rise to a region of
negative $\partial P / \partial T|_V$~\cp{Bagnier2000,Vo07} that is
shown in Fig.~\ref{P_vs_T}. This leads to a negative Gr\"uneisen
parameter and could introduce a barrier to convection in Jupiter and other giant planets. However \ct{MHVTB} demonstrated that in a hydrogen-helium
mixture, the region of $\partial P / \partial T|_V < 0$ is shifted to
lower temperatures than occur in Jupiter.

The question whether the insulator-to-metal transition in hydrogen is
smooth or of first order (plasma phase transition) has been debated
for a long time. A series of free energy models were constructed with
a plasma phase transition~\cp{EF89,SC92}, while others do not include
one~\cp{Ro98,sesame}. Recent shock wave experiments~\cp{Fortov2007}
show evidence of an insulator-to-metal transition at temperature and
pressure conditions that are consistent with theoretical
predictions. In their interpretation, the authors carefully suggest
that their data may provide evidence for a first order phase
transition.

Recently \ct{MHVTB} and independently \cite{NHKFRB} used
first-principles simulations to derive an EOS to model Jupiter's
interior.  While both studies relied on the same simulation technique,
DFT-MD, the groups derived very different predictions for the size of
Jupiter's rocky core, the distribution of heavy elements in its
mantle, and the temperature profile in its interior. The differences
between the two approaches are analyzed in \citet{hedla2008} and will
be summarized later in this chapter. While some deviations in the
simulation parameters such as the number of particles and their effect
on the computed EOS have been identified, the differences in the
predictions are mainly due to additional model assumptions
e.g. whether helium and heavy elements are homogeneously distributed
throughout the interior \cp{MHVTB}, or not \cp{NHKFRB}. Only the
deviations in the interior temperature profile is a direct consequence of
the computed EOS. While pressure and internal energy can be obtained
directly from simulations at constant volume and temperature, the
entropy does not follow directly and is typically obtained by
thermodynamic integration of the free energy. \ct{MHVTB} and
\cite{NHKFRB} used different methods to compute the free energy but
more work is needed to understand which method yields more accurate
adiabats.

Two recent papers~\cp{Morales2009,Lorenzen2009} provide evidence for
the phase separation of the hydrogen-helium mixtures at the interior
of Saturn and possibly also of Jupiter. Both papers rely on DFT-MD
simulations but \ct{Morales2009} used larger and more accurate
simulations and employed thermodynamic integration to determine the
Gibbs free energy of mixing. \ct{Lorenzen2009} used a simplified
approach where mixing entropy, which is the most difficult term to
calculate, was taken from a noninteracting system of particles. While
others have shown that non-ideal mixing effects are
important~\cp{Vo07}, the impact of this approximation on the
pressure-temperature conditions where the hydrogen and helium start to
phase-separate remains to be studied but it may explain why
\ct{Lorenzen2009} predict higher phase-separation temperatures than
\ct{Morales2009}.

In addition to hydrogen and helium, water has also been extensively studied experimentally \cp[e.g.][]{Lee06} and computationally \cp[e.g][]{Cavazzoni99,Schwegler01,French09}.  Often water, ammonia, and methane, are grouped together as ``planetary ices,'' as a generic phrase for O-, C-, and N-dominated volatiles (mostly H$_2$O, NH$_3$, and CH$_4$), which are likely found in fluid form, not solid, within giant planets.  Furthermore, these components are probably not found as intact molecules at high pressure.  Water is found from first-principles calculations to dissociate into H$_3$O$^+$ + OH$^-$ ion pairs above $\sim$2000 K at 0.3 Mbar \cp{Cavazzoni99,Schwegler01,Mattsson06}, and indeed high electrical conductivities are measured at pressures near 1 Mbar \cp{Chau01}.  Moving to even heavier elements, ``rock'' refers primarily to silicates (Mg-, Si- and O-rich compounds) and often includes iron and other ``metals'' as well \cp[see][]{Stevenson85,Hubbard84b}.  Uncertainties in the EOSs of heavy elements are generally less important in structure and evolution calculations than those for hydrogen and helium, but they certainly do have important quantitative effects \cp{Hubbard91,Baraffe08}.

This section is only able to give a flavor of the vast array of EOS science going on at the boundary between physics and planetary sciences.  We are now in an era where long sought-after advances in experiment and first-principles theory are occurring at a fast pace.

\bigskip
\noindent
\textbf{ 2.2 Basic Equations}
\bigskip

To a first approximation, a planet can be considered as a non-rotating, non-magnetic,  fluid object where gravity and gas pressure\footnote{Radiation pressure is negligible in planetary interiors.} are the two main contributors to forces in the interior. Planetary structure and evolution can thus be described
in a spherically symmetric configuration and are governed by the following conservation equations:
\beq
{\rm Mass \, conservation:} \, {\partial m \over \partial r} = 4 \pi r^2 \rho,
\eeq
\beq
{\rm Hydrostatic \,  equilibrium:} \,  {\partial P \over \partial r} = -\rho g,
\eeq
\beq {\rm Energy \, conservation:} \, {\partial L \over \partial r} = -4 \pi r^2 \rho T {\partial S \over \partial t},
\eeq
where $L$ is the intrinsic luminosity, {\it i.e} the net rate of radial energy flowing through a sphere of 
radius $r$.  The rate of change of the matter entropy is due to the variation of its internal energy and to compression or expansion work, according to the first and second laws of thermodynamics. A complete model requires an additional equation describing the transport of energy in the planet:
\beq
{\partial T \over \partial r} = {\partial P \over \partial r} {T \over P} \nabla,
\eeq
where $\nabla = {dlnT \over dlnP}$ is the temperature gradient. 
If energy transport
is due to radiation or conduction, it is well described by a diffusion process with:
\beq
{\partial T \over \partial r} = {3 \over 16 \pi a c G} {\kappa L P \over m T^4}.
\eeq
The total opacity of matter, $\kappa$, accounts for radiative and conductive transport and
is defined by $\kappa^{-1} = \kappa_{\rm Ross}^{-1} +  \kappa_{\rm cond}^{-1} $
with  $\kappa_{\rm Ross}$ and $\kappa_{\rm cond} $ are the Rosseland mean radiative opacity and the conductive opacity, respectively.
Energy transport by conduction in a fluid results from collisions during random motion of particles. For planets essentially composed of H/He, energy transfer is due to electrons in the central ionized part, whereas molecular motion  
dominates in the outer envelope. Because of
the high opacity of H/He matter in planetary interiors, convection is thought to be the main energy 
transport and the temperature gradient is given by
the adiabatic gradient $\nabla_{\rm ad} = ({dlnT \over dlnP})_{\rm S}$ . If the planet has a core made of heavy material (water or ice, rock, iron), heat transport can be due to convection or conduction (electrons or phonons), depending on the core material, its state (solid or liquid) and the age of the planet.

\bigskip
\noindent
\textbf{ 2.3 Thermal Evolution And Atmospheric Boundary Conditions}  \label{evol}
\bigskip

Starting from a high entropy, hot initial state, the luminosity of a planet during its entire evolution is powered  by the release of its gravitational $E_{\rm g}$ and internal $E_{\rm i}$  energy and is given by (see Eq. (3)):
\begin{equation}
L(t)=-{d \over dt} (E_{\rm g} + E_{\rm i}) =- \int_M P {d \over dt} ({1\over \rho})dm\, -\int_M {de \over dt} dm,
\end{equation}
where $e$ is the specific internal energy. The virial theorem, which applies to a self-gravitating gas sphere in hydrostatic equilibrium, relates  the thermal energy of a planet (or star) to its gravitational energy as follows:
\beq
\alpha E_{\rm i} + E_{\rm g} = 0,
\eeq
 with  $\alpha$=2 for a monoatomic ideal gas or a fully non-relativistic 
degenerate gas, and $\alpha = {6 \over 5}$ for an ideal diatomic gas. 
Contributions arising from interactions between particles yield corrections
to the ideal EOS \cp[see][]{Guillot05}.
The case $\alpha = {6 \over 5}$ applies to the molecular hydrogen outer 
regions of a giant planet.
Note that the mass fraction involved in these regions, for a Jupiter-like planet, is 
usually negligible compared to that involved in the central core and the metallic H region.

According to Eqs. (6) and (7), the planet radiates:
\beq
L \propto - {d E_{\rm g} \over dt},
\eeq
defining a characteristic thermal timescale $\tau_{\rm KH}$:
\beq
\tau_{\rm KH} \sim  {  E_{\rm g} \over L}  \sim {G M^2 \over RL}.
\eeq 
For a 1 $\mjup$ gaseous planet, with negligible heavy element content,
$\tau_{\rm KH} \sim 10^7$ yr at the beginning of its evolution \footnote{Note that this value is highly uncertain since it depends on the initial state and thus on the details of the planet formation process.  See \S4.6} and $\tau_{\rm KH} 
> 10^{10}$ yr after 1 Gyr \cp[see e.g.][for values of $R$ and $L$ at a given age]{Baraffe03}.  The reader must keep in mind that Equation (9) is a rough estimate of the characteristic timescale for cooling and contraction of a planet. This timescale can be longer by 1 or 2 orders of magnitude than the value derived from Eq. (9).  Indeed, when degeneracy sets in, a significant fraction of the gravitational energy due to contraction is used to increase the pressure of the (partially) degenerate electrons and the luminosity of the planet is essentially provided by the thermal cooling of the ions \cp[see][]{Guillot05}. 

The rate at which internal heat escapes from a planet depends
on its atmospheric surface properties, and thus on the outer 
boundary conditions connecting inner and atmospheric
structures.  Put another way, although the interior may be efficiently convecting, the radiative atmosphere serves as the bottleneck for planetary cooling.  For objects with cold molecular atmospheres,
the traditional Eddington approximation assuming that 
the effective temperature equals
the local temperature at an optical depth $\tau_{\rm Ross}=2/3$ 
provides incorrect thermal profiles and large errors on $\te$ 
\cp[see][and references therein]{Chabrier00}.
Modern models for planets incorporate 
more realistic atmospheric boundary conditions using frequency dependent
atmosphere codes. Inner and outer temperature-pressure profiles must be
connected
at depths where the atmosphere becomes fully convective, implying an
adiabatic thermal profile, and optical depth is greater than one.
The connection is done at a  fixed pressure, usually a few bars \cp{Burrows97,Burrows03,FH03,Guillot05} or at a fixed optical depth, usually at
$\tau=100$ \cp{Chabrier00}. The numerical 
radius corresponding to the outer boundary conditions provides, 
to an excellent approximation, the planet's photospheric radius, 
where the bulk of the flux escapes.

\bigskip
\noindent
\textbf{ 2.4 Effect Of Rotation}
\bigskip

The solar system giant planets are relatively fast rotators with 
periods of about 10 hours for Jupiter and Saturn, and about 17 hours for Neptune and Uranus \cp[see][and references therein]{Guillot05}. Rotation modifies the internal structure of a fluid body and yields departures from a spherically symmetric configuration. Its effects can  be accounted for using a perturbation theory, which has been 
extensively developed during the past century for planets and stars \cp{Tassoul78,ZT74,ZT76,ZT78}. 
An abundant literature exists on the application of this theory to models for  our 
solar system giant planets \cp{Hubbard89,Guillot94b,Podolak95}. One often refers to the so-called theory 
of figures, presented in details in the works of {\em Zharkov and Trubitsyn}.

Here we only briefly explain the concept of the theory. The underlying idea
 is to define a rotational potential $V_{\rm rot}$ such that surfaces of constant $P$, $\rho$ and $U$ coincide. $U = V_{\rm rot} + V_{\rm grav}$  is the total potential (rotational + gravitational) and obeys the hydrostatic equilibrium equation $\nabla P = \rho \nabla U$. The solution for the level surfaces (or figures) can be expressed in terms of an expansion in even Legendre polynomials, also called zonal harmonics, $P_{\rm 2n}$. As a consequence of rotation,
the gravitational potential departs from a spherically symmetric  potential and can be expressed in terms of even Legendre polynomials and gravitational moments 
$J_{\rm 2n}$ (also called zonal gravitational harmonics) as follows:
\beq
V_{\rm grav} = {GM \over r} \left [1 - \sum_{\rm n=1}^{\infty} ({a \over r})^{\rm 2n} J_{\rm 2n} P_{\rm 2n}(cos \theta) \right ],
\eeq
where $r$ is the distance to the planet center, $a$ its equatorial radius and $\theta$ the polar angle to the rotation axis. The gravitational moments can be related to the density profile of the planet (see {\em Zharkov \& Trubitsyn}, 1974) as follows:
\beq
J_{\rm 2n} = -{1 \over Ma^{\rm 2n}} \int\int\int \rho(r,\theta)r^{\rm 2n}P_{\rm 2n}(cos \theta) dv,
\eeq
where $dv$ is a volume element. A measure of $J_{2n}$ by a spacecraft coming close to a planet thus provides constraints on its density profile. This powerful and elegant method was applied to our four giant planets using the determination of $J_2$, $J_4$ and $J_6$ from the trajectories of the space missions Pioneer and Voyager 
(see {\em Guillot}, 2005, for a review, and the section below, ``Solar System Giant Planets'').

As is well known from the theory of figures, the low order gravitational moments provide constraints on the density/pressure profile in the metallic/molecular hydrogen regions but do not sound the most central regions \cp[][see also Fig.~4 of {\em Guillot}, 2005]{Hubbard99b}. Consequently, the presence of a central core, its mass, composition and structure, can only be {\em indirectly} inferred from the constraints  on the envelope provided by the gravitational moments. 

\begin{figure}
 \epsscale{1}
 \plotone{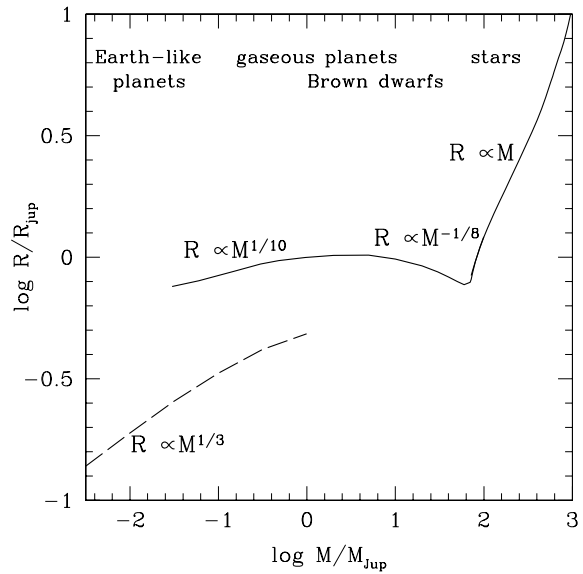}
  \caption{\small Characteristic mass-radius relationship for stellar 
and substellar gaseous H/He objects (solid line, models from {\em Baraffe et al},
1998, 2003, 2008) and for pure ice planets (dashed line, models from
{\em Fortney et al.} 2007).}  \label{figmr}
 \end{figure}

\bigskip
\noindent
\textbf{ 2.5 Mass-Radius Relation}  \label{mrtheory}
\bigskip

The mass-radius relationship for planets, and more generally for substellar/stellar 
objects, contains essential informations about their main composition and 
the state of matter in their interior. The fundamental work by \ct{Zapolsky69} is a perfect illustration of this statement. 
The analysis of cold (zero-temperature) spherical bodies of a given chemical composition
and in hydrostatic equilibrium shows the existence of a unique mass-radius relation
and of a maximum radius $R_{\rm max}$ at a critical mass $M_{\rm crit}$.  The very 
existence of a maximum radius stems from two competing physical effects characteristic of the
state of matter under planetary conditions. The first effect is due to electron 
degeneracy, which dominates at large masses and yields a 
mass-radius relationship $R \propto M^{-1/3}$ characteristic of 
fully degenerate bodies \cp{Chandra39}. 
The second effect stems from the classical electrostatic contribution
from ions (Coulomb effects) which yields a mass radius relation 
$R \propto M^{1/3}$, characteristic of incompressible Earth-like planets. 
\ct{Zapolsky69} find a
critical mass of 2.6 $\mjup$ where the radius reaches a maximum value 
 $R_{\rm max} \sim$1 $\rjup$ for a gaseous H/He planet. The critical mass
increases as the heavy element content increases, while  $R_{\rm max}$ decreases. 
 
The true mass-radius relationship, derived from models taking 
into account a realistic equation of state (see \S 2.1)
yields a smoother dependence of 
radius with mass, as displayed in Fig. \ref{figmr}. The transition between
stars and brown dwarfs marks  the onset of electron 
degeneracy, which inhibits the stabilizing generation of nuclear energy
by hydrogen burning. The typical transition mass
is $\sim$ 0.07 $\msol$ \cp{Burrows01,Chabrier00}. 
Above this transition mass, the nearly classical ideal gas
yields a mass-radius relationship $R \propto M$. In the brown dwarf regime
the dominant contribution of partially degenerate electrons, balanced by
the contribution from ion interactions yields $R \propto M^{-1/8}$ 
instead of the steeper 
relationship for fully degenerate objects. The increasing
contribution of Coulomb effects as mass decreases competes with electron
degeneracy effects and renders the radius almost
constant with mass around the critical mass. The full calculation
yields, for gaseous H/He planets,  $M_{\rm crit} \sim 3 \mjup$,
amazingly close to the results based on the simplified approach
of \ct{Zapolsky69}. Below the critical mass,
Coulomb effects slightly dominates over partially degenerate effects, 
yielding a smooth variation of radius with mass close to the relation 
$R \propto M^{1/10}$.

\bigskip
\section{SOLAR SYSTEM GIANT PLANETS} \label{ss}
\bigskip
\noindent
\textbf{ 3.1 Jupiter And Saturn}
\bigskip

Jupiter and Saturn, composed primarily of hydrogen and helium, serve as the benchmark planets for our understanding of gas giants.  Indeed, these planets serve as calibrators for the structure and cooling theory used for all giant planets and brown dwarfs.  The significant strides that have been made in understanding these planets have come from theoretical and experimental work on the equation of state of hydrogen, as well at the spacecraft-measured gravitational moments.

As discussed in \S \ref{mrtheory}, with a basic knowledge of the EOS of hydrogen, helium, and heavier elements, along with observations of Jupiter's mass and radius, one can deduce that Jupiter and Saturn are mostly composed of hydrogen \cp[see, e.g.][]{Seager07}.  The same argument shows that Uranus and Neptune are mostly composed of elements much heavier than hydrogen.

Models of the interior structure of Jupiter in particular were investigated by many authors in the 20th century, and investigations of Jupiter and Saturn became more frequent with the rise of modern planetary science in the 1950s and 1960s.  However, even these initial models were uncertain as to whether the hydrogen is solid or fluid.  The path towards our current understanding of giant planets started with the observation by \ct{Low66} that Jupiter emits more mid infrared flux than it received from the Sun.  Soon after, \ct{Hubbard68} showed that the observed heat flux could only be carried by convection (as opposed in radiation or conduction), showing that Jupiter's interior is fluid, not solid.  Thus began the paradigm of giant planets as hydrogen dominated, warm, fluid, convective objects.

However, it was clear that Jupiter and Saturn were not composed entirely of hydrogen and helium in solar proportions---they have radii too small for their mass.  \ct{Podolak74} showed that these planets likely had massive heavy elements cores, which tied their interior structure to a possible formation mechanism \cp{Perri74,Mizuno78}.  \ct{SS77b,SS77a} performed detailed investigations of phase diagrams and transport properties of fluid hydrogen and helium, putting our understanding of these planets on a firmer theoretical footing.  The issues discussed in these papers still reward detailed study.  Also in the early to mid 1970s, the first thermal evolution models of Jupiter were computed \cp{Grossman72,Bodenheimer74,Graboske75} and connections were made between these models and similar kinds of cooling models for fully convective very low mass stars.  Much of the early knowledge of giant planet structure and evolution is found in \ct{Hubbard73b} and \ct{ZT78}.  A consistent finding over the past few decades is that fully adiabatic cooling models of Jupiter can reproduce its current \teff\ to within a few K.

\begin{figure}
 \epsscale{1.0}
 \plotone{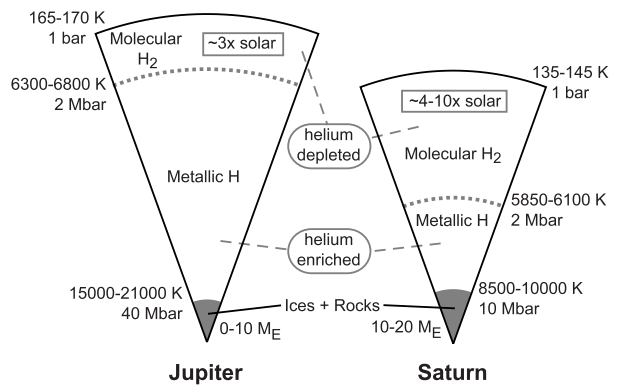}
  \caption{\small Interior views of Jupiter and Saturn, from calculations with the \ct{SCVH} ``chemical picture'' H/He EOS.  Jupiter is more massive, which leads to a greater fraction of its mass in the high-pressure liquid metallic phase.  It also has a higher temperature at a given pressure.  Interior temperatures are taken from \ct{Guillot05}.
}  \label{js}
 \end{figure}

The situation for Saturn is more complicated.  There is a long-standing cooling shortfall in thermal evolution models of Saturn:  the planet is $\sim$50\% more luminous than one calculates for a 4.5 Gyr-old, adiabatic, well-mixed planet \cp{Pollack77,SS77b}.  The most likely explanation is that the He is currently phase separating from the liquid metallic hydrogen \cp{Stevenson75}, and has been for the past 2-2.5 Gyr.  This immiscible He should coalesce to form droplets, that are denser than the surrounding H/He mixture, and then ``rain'' down within the planet.  This differentiation is a change of gravitational potential energy into thermal energy.  Recent models of the evolution of Saturn with the additional energy source indicate that the He may be raining down on top of the core \cp{FH03}.  The largest uncertainly in properly including He phase separation into cooling models is the current observed abundance of He is Saturn's atmosphere, which does appear to be depleted in He relative to protosolar abundances, but the error bars are large \cp{CG00}.\footnote{Note that this ``He rain'' cannot power the inflated hot Jupiters, as their large radii require interior temperatures warmer than needed for phase separation.}  The \emph{in situ} observation that Jupiter's atmosphere is modestly depleted in helium as well shows that our understanding of the evolution of both these planets is not yet complete.

Figure \ref{js} shows models of Jupiter's and Saturn's interiors taken from \ct{Guillot05}.  As Jupiter is 3.3 times more massive than Saturn, a greater fraction of its interior mass is found at high pressure.  Therefore, most of the hydrogen mass of Jupiter is fluid metallic, while for Saturn it is fluid molecular, H$_2$.  Jupiter's visible atmosphere is warmer than Saturn's.  If their interiors are fully adiabatic, this implies that Jupiter is always warmer at a given pressure than Saturn.  Since Jupiter has a larger interior heat content (residual energy leftover from formation), this also means that at a given age, Jupiter is always more luminous than Saturn.

Structural models for Saturn show that it is more centrally condensed than Jupiter.  Although significant central condensation is expected due to the compressibility of H/He, detailed models of Saturn indicate even greater central condensation is needed to match its gravitational field.  This confirms that a significant fraction of the heavy elements must be in the form of central dense core.  Recent estimates from \ct{Saumon04} indicate a core mass of 10-20 \me.  The majority of Saturn's heavy elements are within the core.  For Jupiter, the situation is less clear cut for several reasons.  First, a 10 \me\ core would only be 3\% of the planet's total mass, so EOS inputs must be accurate to this same percentage for real constraints on the core mass.  Second, since Jupiter is more massive than Saturn, a greater fraction of its interior is in the 1-100 Mbar region that it difficult to model.  \ct{Guillot99} and \ct{Saumon04} found that Jupiter models without any core were allowed, and that the majority of the planet's heavy elements must be mixed into the H/He envelope.  This would be a clear difference compared to Saturn.  On the whole Saturn possesses $<25$ \me\ of heavy elements, while Jupiter is $<40$ \me.  Saturn appears to be a relatively greater fraction of heavy elements by mass.

Very recently, two groups have computed new models of the interior of
Jupiter, based on first-principles equations of state for hydrogen and
helium. \ct{MHVTB} simulate the hydrogen-helium mixtures directly,
under the pressure-temperature conditions found in the interior of
Jupiter. This work predicts a large core of 14 -- 18 \me\ for Jupiter,
which is in line with estimates for Saturn and suggests that both
planets may have formed by core-accretion. The paper further predicts
a small fraction of planetary ices in Jupiter's envelope, suggesting
that the ices were incorporated into the core during formation rather
than accreted along with the gas envelope.  Jupiter is predicted to
have an isentropic and fully convective envelope that is of constant
chemical composition. In order to match the observed gravitational
moment $J_4$, the authors suggest that Jupiter may not rotate as a
solid body and predicted the existence of deep winds in the interior
leading to differential rotation on cylinders. Differential rotation
as well as the size of Jupiter's core can potentially be measured by
the forthcoming
\emph{Juno} orbiter mission, briefly described below.

\begin{figure}
 \epsscale{1.0}
 \plotone{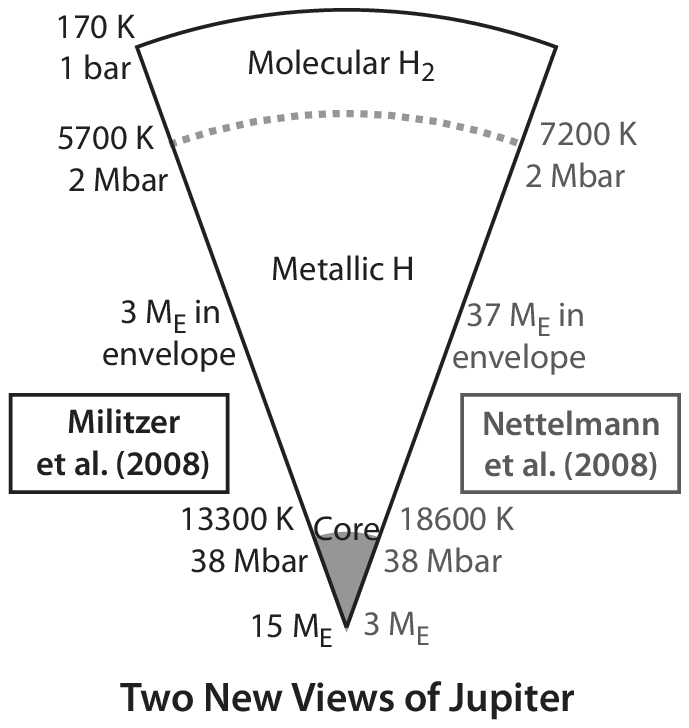}
  \caption{\small Possible revised interior views of Jupiter by \ct{MHVTB} and \ct{nettelmann08}.  Temperature, pressures, and abundances of heavy elements are labeled.
}  \label{newj}
 \end{figure}

Alternatively, \ct{nettelmann08} computed EOSs for hydrogen, helium, and water separately, and investigate interior models under the assumption of an additive volume rule at constant temperature and pressure.  They hypothesize a heavy element-enriched metallic region and heavy element-depleted molecular region.  They overall find a large abundance of heavy elements within Jupiter, but far less in a distinct core.  A further key difference between these two models that has not yet been investigated is interior temperatures--the \ct{nettelmann08} model predicts higher temperatures than do \ct{MHVTB}.  This will have important consequences for cooling models of the planet, which will need to be investigated. The \ct{nettelmann08} and the \ct{MHVTB} approaches are compared in \citet{hedla2008}, while Figure \ref{newj} gives a schematic comparison.

The resolution of the H/He phase separation issue in Saturn may only come from an entry probe into Saturn's atmosphere, which could measure the He/H$_2$ ratio \emph{in situ}, as was done for Jupiter.  If the He mixing ratios were known precisely in each atmosphere, it is possible that phase diagram and evolution history combination could be derived that constrains the demixing region of the high-pressure H/He phase diagram \cp{FH03}.  In theory, if atmospheric abundances of He, C, N, and O were known, this would be a strong constraint on the properties of the entire H/He envelope.  In practice, this has not yet happened for Jupiter because the \emph{Galileo Entry Probe} was apparently only able to measure a lower limit on the O abundance from water vapor.  The forthcoming \emph{Juno} mission will use microwave spectroscopy to measure the deep H$_2$O and NH$_3$ abundances at pressures near 100 bars (below cloud condensation levels), which should help to resolve this issue for Jupiter.  \emph{Juno} is expected to launch in 2011 and arrive at Jupiter in 2016, with detailed data analysis beginning in 2018.

\emph{Juno}, which is a low periapse orbiter, will also exquisitely map the planet's gravitational and magnetic fields.  Work by \ct{Hubbard99b} has shown that if surface zonal flows extend down to 1000 km depth ($P\sim10$ kbar) then this should be observable in the planet's gravity field.  This will give us a view of the \emph{mechanics} of the interior of the planet.  Does it rotate as a solid body, on cylinders, or something more exotic?  Furthermore, the tidal response of Jupiter to the Galilean satellites may also be detectable, which will help to constrain the planet's core mass.

The \emph{Juno} orbiter will also map Jupiter's magnetic field.  The magnetic fields of Jupiter and Saturn are large and predominantly dipolar, with a small tilt from the planet's rotation axis, similar to the Earth.  This tilt is 9.6$^{\circ}$ for Jupiter, but less then 1$^{\circ}$ for Saturn.  These fields are consistent with their production via a dynamo mechanism within the liquid metallic region of the interior.  Unfortunately, dynamo physics is not understood well enough to place strong constraints on the interiors of Jupiter and Saturn.

\bigskip
\noindent
\textbf{ 3.2 Uranus And Neptune}
\bigskip

Uranus and Neptune have not received the same attention that have been paid to Jupiter and Saturn.  The neglect of the ``ice giants'' has been due to relatively less precise observational data, and the complicated picture that the data has revealed.  The first stumbling block has been the relatively high density of these planets, compared to Jupiter and Saturn.  This high density shows that these planets are \emph{not} predominantly composed of hydrogen and helium.  But then what is their composition?  Mostly the fluid planetary ices of water, ammonia, and methane?  Mostly rock and iron?  The mass/radius of these planets can be matched with a very wide range of compositions from the three categories of H/He gas, ice, and rock \cp[e.g.][]{Podolak95}.  A high pressure mixture of rock and H/He can very nicely mimic the pressure-density relation of the ices \cp{Ubook}, which means that even gravity field data, which helps to elucidate central condensation, cannot break the ice vs.~rock/gas degeneracy.  Therefore, modelers often have had to resort to cosmogonical constraints, such as an assumed ice-to-rock ratio from protosolar abundance arguments, to constrain structural models.  Recent estimates of the H/He mass fraction yield 1-2 \me\ in both planets (with an ice-to-rock ratio of 2.5), with a hard upper limit of $\sim$5 \me\ in each if only rock and H/He gas are present \cp{Podolak00}.

Nearly all of collected knowledge on the structure and evolution of these planets have been gathered in the \emph{Uranus} and \emph{Neptune and Triton} Arizona Space Science Series book chapters by \ct{Ubook} and \ct{Nbook}, respectively.  These chapters summarize our state of knowledge of Uranus and Neptune as of the early 1990s, after the Voyager 2 encounters.  Importantly, novel research is also presented in those chapters that is not found in the more readily available literature.  Both chapters are worth detailed study.  Since that time, the only more recent update is the work of Marley and Podolak, who investigated Monte Carlo interior models of these planets with a minimum of assumptions regarding interior density \cp{Marley95,Podolak00}.  Without assumptions regarding the layering of gas, ices, and rock, these authors perform Monte Carlo studies of the interior density distribution, which uses the gravity field alone to show where density jumps, if any, must occur.

\begin{figure}
 \epsscale{1.0}
 \plotone{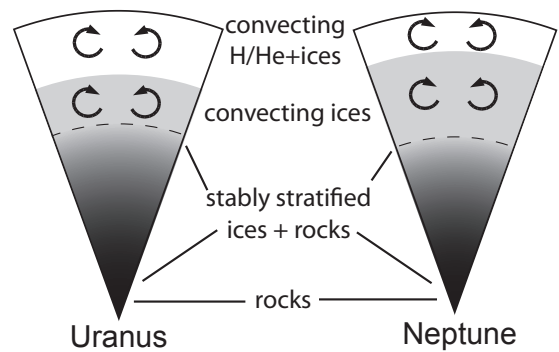}
  \caption{\small Possible interior views of Uranus and Neptune.  White indicates a composition of predominantly H$_2$/He gas (with smaller amounts of heavy elements mixed in), solid gray is predominantly ices, and black predominantly rock.  The gray-to-black gradient region in each planet shows where the interior may be statically stable due to composition gradients.  Circles with arrow heads indicate convection.  Neptune appears to be composed of a greater fraction of heavy elements and it may have a larger freely convective region.  This and other inferences are, however, uncertain.}
  \label{un}
 \end{figure}

The second major complication with these planets, after composition degeneracy, is the interior heat flow.  While fully adiabatic, fully convective thermal evolution models reproduce the current luminosity of Jupiter, and underpredict the luminosity of Saturn, they \emph{overpredict} the luminosity of Neptune and Uranus.  The situation for Uranus is especially dramatic, as no intrinsic flux from the planet's interior was detected by Voyager 2.  At least two important ideas partially address the heat flow issue.  \ct{Hubbard80b} suggested that the absorbed and reradiated stellar flux may be large enough to swamp the intrinsic flux, a smaller component.  This would be a larger effect in Uranus than Neptune, since it is closer to the Sun.  This same effect, on a much more dramatic scale, is seen for the hot Jupiters, where the intrinsic flux is unmeasurable, since it is $10^4$ smaller than reradiated absorbed flux.  While this effect is certainly real in Uranus and Neptune, it alone cannot explain the low heat flows \cp{Nbook}.

The problem may well be in the assumption that the interior is partitioned into well defined layers of H/He gas, the fluid ices, and rock.  If these distinct layers exist, then convection should be efficient in each layer, and the interior heat should be readily transported to the surface.  However, it is well known that composition gradients can readily suppress convection, as a much steeper temperature gradient is needed for convective instability to occur, from the Ledoux criterion.  If large regions of the interior of these planets are stably stratified, then stored residual energy from formation will be ``locked'' into the deep interior, and will only be transported quite slowly.  At gigayear ages this would lead to a small intrinsic luminosity.  A promising explanation for the reduced heat flow of Uranus and Neptune is that the deep interiors of the planets, which are likely a mix of fluid ices and solid rock, are predominantly stratified, with only the outer $\sim$1/3 of the heavy element interior region freely convecting \cp{Nbook}. 

Recently, interior geometry as proposed above was investigated with 3D numerical dynamo models \cp{Stanley06}.  Uranus and Neptune are known to have complex magnetic fields that are non-dipolar,  non-axisymmetric, and tilted significantly from their rotation axes.  Stanley \& Bloxham have found that they can reproduce the major features of these fields with an outer convecting ionic shell, in the outer $\sim$20-40\% of the interior region, with the innermost regions not contributing to the dynamo.  This is strong evidence that the interiors of these planets are complex and are not fully convective.  In addition, structure models that include a pure H/He envelope, a pure icy layer, and an inner rocky core are not consistent with the gravity field data of either planet.  A possible interior view of each planet is shown in Figure \ref{un}.

Since we will not have additional data to constrain the gravitational moments of these planets, or an entry probe, for perhaps decades, further progress in understanding the structure and evolution of Uranus and Neptune must come from new theoretical ideas.  At this time, new thermal evolution models are needed to explore in some detail what regions of the interior are indeed convective.  Future work on heat transport in double-diffusive convective regions in planetary interiors (see \S \ref{anom}) should focus on Uranus and Neptune as well as exoplanets.

\bigskip
\bigskip
\section{RECENT HIGHLIGHTS:  TRANSITING AND\\DIRECTLY IMAGED PLANETS}

\bigskip
\noindent
\textbf{4.1 The Observed Mass-Radius Relationship}
\bigskip

\begin{figure}
 \epsscale{1}
 \plotone{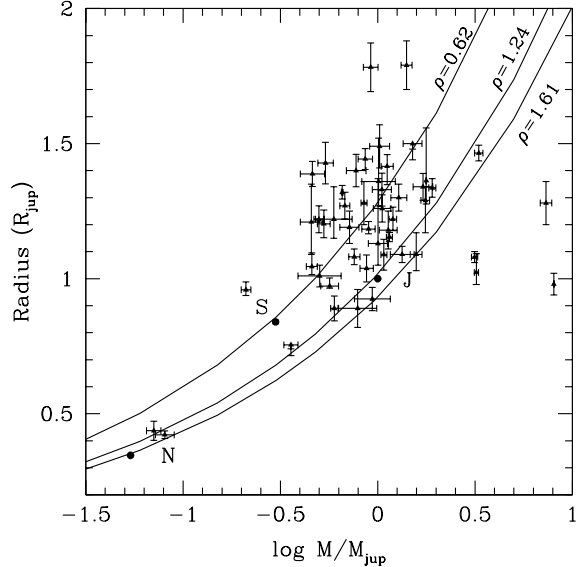}
  \caption{\small Mass-radius diagram for the known transiting planets
  (data are taken from the web site of Frederic Pont: www.inscience.ch/transits).
Three iso-density curves are shown for the mean densities of Saturn 
($\bar{\rho}=0.62$),
Jupiter ($\bar{\rho}=1.24$) and Neptune ($\bar{\rho}=1.61$). 
These three solar system giant planets are also indicated by solid points.}  \label{figtransit}
 \end{figure}

The  determination of mass-radius relationships of exoplanets,
using photometric transit and Doppler follow-up techniques,
provides an unprecedented opportunity to extend our 
knowledge on planetary structure and composition. The 20th century
was marked by the solar system 
exploration by space missions, revealing the 
complexity and diversity of giant planets in 
terms of  chemical composition and internal structure.
The prolific fishing for transiting exoplanets in the beginning of this 21st century, with a catch of  over fifty objects, 
not only confirms this diversity but also raises new questions
in the field of planetary science. Two benchmark discoveries illustrate
the surprises planet hunters were faced with. The very first transiting
planet ever discovered, HD 209458b \cp{Charb00,Henry00}, 
 was found with an abnormally large radius, a puzzling property now shared by a 
growing fraction of transiting exoplanets. At the other extreme,
 a Saturn mass planet, HD 149026b \cp{Sato05} was discovered 
with such a small radius that more than 70 $\mearth$ of heavy elements 
is required to explain its compact structure. This discovery raised in particular
new questions on the formation process of planets with such a large amount 
of heavy material. The diversity in mean density of transiting planets yet discovered is illustrated in 
Fig. \ref{figtransit}. 

\bigskip
\noindent
\textbf{4.2 Irradiation Effects}
\bigskip

The discovery of HD 209458b and of the additional transiting exoplanets that followed has opened
a new era in giant planet modeling. The modern theory of exoplanet radii 
 starts with models including irradiation effects from the 
parent star. These effects on planet evolution are accounted for
through the coupling between inner structure models and irradiated
atmosphere models, following the same method described in \S 2.3.
Current treatments are based on simplified treatments of the atmosphere, using 1D plane-parallel
atmosphere codes. They however allow one
to understand the main effects on planetary evolution. 
In reality the impinging stellar flux has an angle of incidence which is a
 function of the latitude and longitude, but in 1D one attempts to compute a planet-wide or day-side average atmosphere profile, using a parameter $f$ which represents the redistribution factor of the 
stellar flux over the planet surface \cp{Baraffe03,Burrows03,Fortney06}. 
The incident stellar flux $F_{\rm inc}$ is explicitly included 
in the solution of the radiative transfer equation and in the 
computation of the atmospheric structure and is defined by:
\beq
F_{\rm inc} = {f \over 4} \, \left( {R_* \over a} \right) ^2 F_*,
\eeq
where $R_*$ and $F_*$ are the stellar radius and flux respectively,
and $a$ the orbital separation.
The current generation of models often use $f$=1, corresponding to a 
stellar flux redistributed  over the entire planet's surface 
or $f=2$ if heat is redistributed only over the day side.
Heat redistribution is a complex problem of atmospheric dynamics, depending
in particular on the efficiency of winds to redistribute energy from
the day side to the night side. 
This question is a challenge for atmospheric circulation
modelers \cp[Showman et al.~chapter, this volume, as well as ][]{Showman08b}. This nascent field 
is growing rapidly with
observational constraints provided by infrared lightcurves
obtained with {\em Spitzer}, which are starting to 
provide information on the temperature structure, composition and 
dynamics of exoplanet atmospheres.

\begin{figure}
 \epsscale{1}
 \plotone{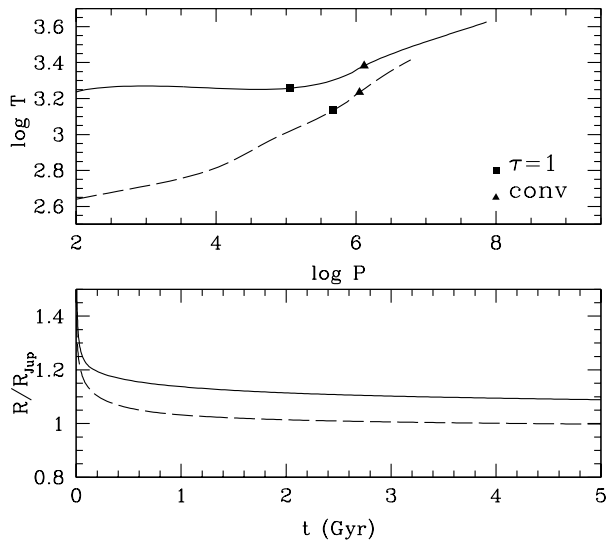}
  \caption{\small Effect of irradiation for a planet at 0.05 AU from the Sun. {\em Upper panel:}
  $T-P$ profiles of atmosphere models with intrinsic $\te$=1000 K (a high value representative of giant planets at young ages) and surface gravity 
  $\log g$=3 (cgs). The solid line corresponds to the irradiated model and the dashed line
  to the non-irradiated model. The locations of the photosphere ($\tau \sim 1$) and
  of the top of the convective zone are indicated by symbols (Models after {\em Barman et al.}, 2001).
  {\em Lower panel:} evolution of the radius with time of a 1 $\mjup$ giant planet.
  Irradiated case: solid line; Non irradiated case: dashed line (Models after {\em Baraffe et al.}, 2003).
 }  \label{figirrad}
 \end{figure}

Although treatments of irradiation effects can differ in the details, with possible refinements accounting for 
phase and angle dependences of the incident flux  
\cp{Barman05,Fortney07b},  different models converge
toward the same effect on the planet atmosphere and evolutionary properties. Atmospheric thermal profiles are strongly
modified by irradiation effects \cp[see Burrows chapter, this volume, as well as ][]{Barman01,Sudar03} as illustrated in Fig. \ref{figirrad} (upper panel).
The heating of the outer layers by the incident stellar flux
yields to an isothermal layer between the top of the convective zone
and the region where the stellar flux is absorbed. The top of the convective
zone is displaced toward larger depths, compared to the non-irradiated
case. The main effect of the shallower atmospheric pressure-temperature profile is to drastically reduce the heat 
loss from the planet's
interior, which can maintain higher entropy for a longer time \cp{Guillot96}.
Consequently, the gravitational contraction of an irradiated planet
is slowed down compared to the non-irradiated counterpart and the upshot
is a larger radius at a given age (see Fig. \ref{figirrad}, lower panel)

The quantitative effect on the planet's radius
depends on the planetary mass, parent star properties and orbital distance. 
Typical effects on the radius of irradiated 
Saturn mass or Jupiter mass planet located at orbital distances
ranging between 0.02 AU and 0.05 AU around a solar-type star are of the order of 
10\%-20\% \cp{Baraffe03,Burrows03,Chabrier04,Arras06,Fortney07a}. 
A consistent comparison between  the theoretical radius
and the observed transit radius requires an additional
effect due to the thickness of the planet atmosphere \cp{Baraffe03,Burrows03,Burrows07}. The measured radius
is a transit radius at a given wavelength, usually in the optical, where the \emph{slant} optical depth reaches $\sim$1.  This is at 
atmospheric layers above the photosphere \cp{Hubbard01,Burrows03}. The latter region is defined by an averaged normal 
optical depth $\tau \sim$1, where the bulk of the flux is emitted outward and which corresponds to the location of the theoretical radius.  The atmospheric extension due to the heating of the incident stellar flux
can be significant, yielding a measured radius larger than the simple
theoretical radius.
This effect can add a few \% (up to 10\%)
 to the measured radius \cp{Baraffe03,Burrows03}. Effects of irradiation on both
the thermal atmosphere profile and the measured radius must be included  for a detailed comparison with observations and can explain some of the less inflated exoplanets.
They are however insufficient to explain the largest radii of
currently known transiting exoplanets, such as TrES-4 with a radius $R_{\rm} = 
 1.78 \rjup$ \cp{Sozzetti08}. This fact
points to other mechanisms to inflate close-in planets.

\bigskip
\noindent
\textbf{4.3 Determining Transiting Planet Composition}
\bigskip

The fraction of planets that are larger than can easily be explained \cp[][put his fraction at $\sim$40\%]{Miller09} serve as a handicap to the goal to understand the composition of these planets.  We would like to be able to constrain the fraction of planetary mass that is the non-H/He heavy elements, and understand how this varies with planetary mass, stellar mass, stellar metallicity, distance from the parent star, and other factors.  Work on understanding the heavy element enrichment of planets has progressed, but must be regarded as incomplete, due to the poorly understood large-radius planets.

It is certainly clear that a number of exoplanets, like our solar system giant planets
(see \S \ref{ss}), are enriched in heavy material, as these planets have radii smaller than pure H/He objects. 
This idea is supported by our current understanding of planet formation via
the core-accretion scenario. This model is itself
consistent with  current observations showing that metal rich environments
characterized by the high metallicity of the parent star 
favors planet formation \cp{Udry07}.

\begin{figure}
 \epsscale{1.0}
 \plotone{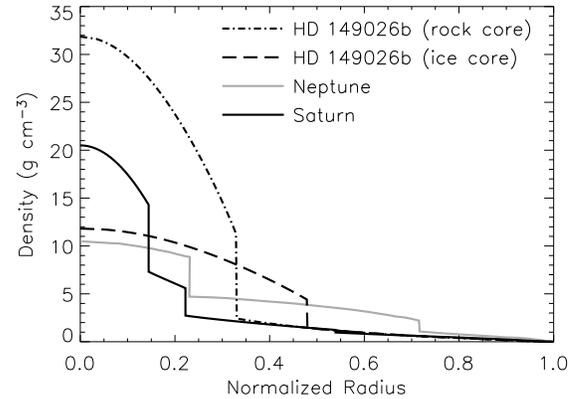}
  \caption{\small Interior density as a function of normalized radius for two possible models for \hh~compared with Neptune and Saturn.  All planet models have been normalized to the radius at which $P$=1 bar.  The Neptune profile is from \citet{Podolak95} and the Saturn profile is from \citet{Guillot99}.  The Saturn and Neptune models have a two-layer core of ice overlying rock, but this is only an assumption.  For Neptune in particular the interior density profile is uncertain.  The two profiles of \hh~assume a metallicity of 3 times solar in the H/He envelope and a core made entirely of either ice or rock.  (After \emph{Fortney et al.}~2006.)
}  \label{intprof}
 \end{figure}

\begin{figure}
 \epsscale{0.9}
 \plotone{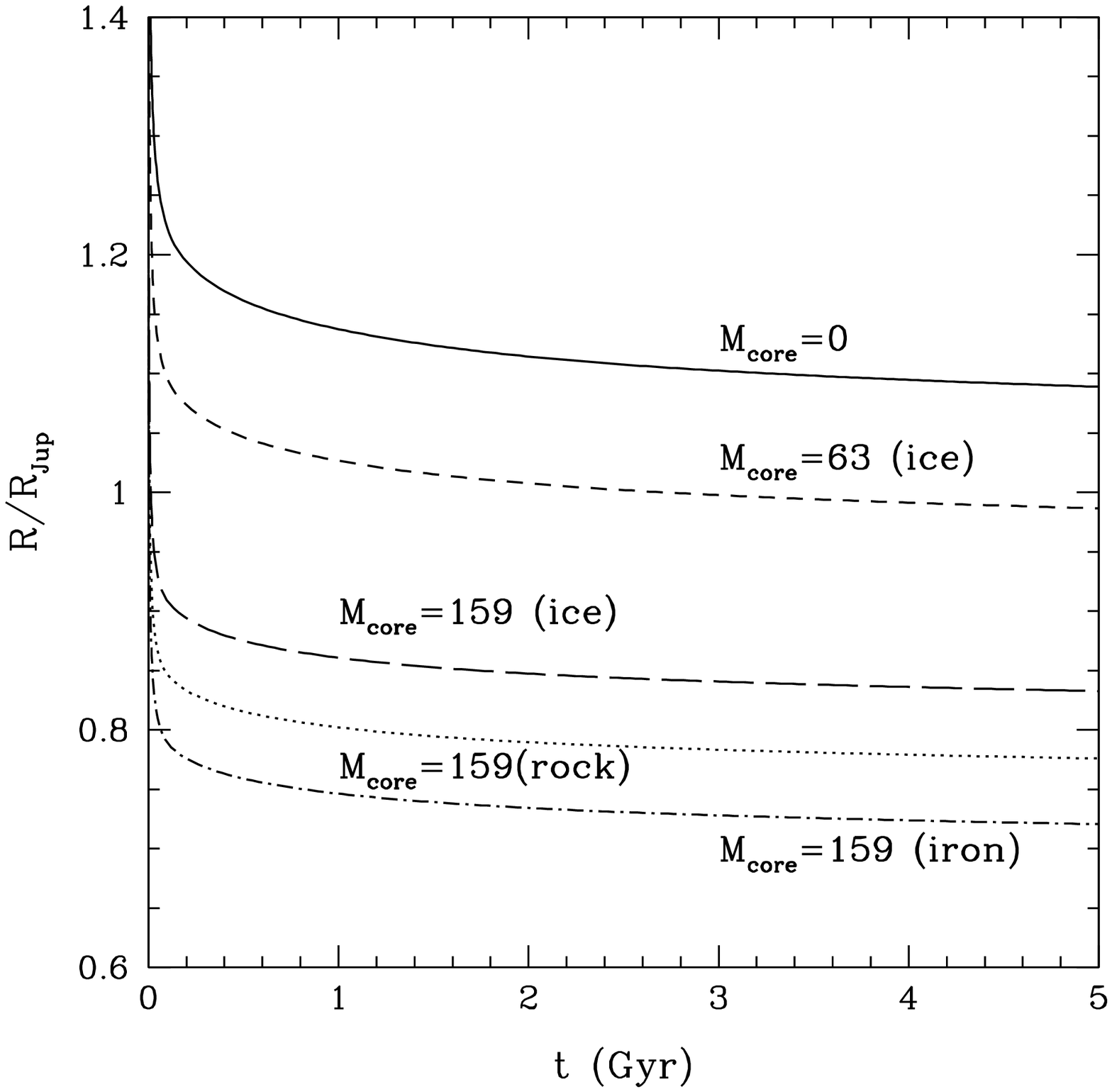}
  \caption{\small Effect of heavy element enrichment on the 
evolution of the radius with time  of a 1 $\mjup$ giant planet.
Solid line: H/He planet; dashed line: 
$\mcore= 63 \mearth$ (ice); long dashed line: $\mcore= 159 \mearth$ (ice);
 dotted line: $\mcore= 159 \mearth$ (rock); dash-dotted 
line: $\mcore= 159 \mearth$ (iron).
}  \label{figtr}
 \end{figure}

The exoplanets with the largest fraction of mass in heavy elements known at the time
this chapter is written are
WASP-7b, with $M_{\rm p} = 0.96 \mjup$ and $R_{\rm p}=0.915 \rjup$
\cp{Hellier08}, HAT-P-3b with $M_{\rm p} = 0.6 \mjup$ and 
$R_{\rm p}= 0.89\rjup$ \cp{Torres07} and 
HD 149026b with $M_{\rm p} = 0. 36 \mjup$ and $R_{\rm p}=0.755 \rjup$ 
\cp{Sato05}. Their small radii
indicate a global mass fraction of heavy material greater than
what Jupiter contains (more than 12\%) and in the case of HD 149026b,
significantly greater than what Saturn contains (more than 30\%).  This 
suggests that the presence of a significant amount of 
heavy material is a property shared by all planets.
Evolutionary models must thus take into account such enrichment and many efforts are
currently devoted to the construction of a wide range of models 
with different amounts  of heavy elements. Exploration of the effects of materials of different
composition is limited by the available equations of state under the conditions
of temperature and pressure characteristic of giant planets.  As a simple starting point, and given the large uncertainty
on the nature of heavy elements and their 
distribution inside planets, current models often
assume that all heavy elements are located in
the core and are water/ice, rock and/or iron.  A possible interior density profile of HD 149026b with a core of either ice or rock is shown in Figure \ref{intprof}.  Uncertainties in current 
planetary models due to uncertainties in the available EOS, the distribution
of heavy elements, and their chemical composition have been analyzed
in detail in \ct{Baraffe08}.
Depending on the total amount of heavy material and its composition,
the radius of an enriched planet can be significantly smaller compared
to that of H/He dominated giant planets. This is illustrated in Fig. \ref{figtr},
which compares the radius evolution with time of planets with different
core sizes (in $\mearth$) and different core compositions.

As noted, the problematic planets are those that are ``too large,'' such that we cannot constrain the heavy element mass in their interiors.  However, if the ``true'' additional energy source or contraction-stalling mechanism affects nearly all hot Jupiters to some degree (which seems likely given that $1/3$ to $1/2$ of planets are too large), then we may be able to take out its effect on planetary radii, and examine the heavy element enrichments of the transiting planets as a collection.  This was first done by \ct{Guillot06} who postulated an unspecified additional interior energy source equal to 0.5\% of the absorbed incident energy.  All planets then required at least some amount of heavy elements to match the measured radii, and they found a correlation between stellar metallicity and planetary heavy element mass.  Subsequently, \ct{Burrows07} found a similar correlation, using only enhanced atmospheric opacities for all planets, although this cannot explain the largest-radius planets. Importantly, \ct{Guillot08} find that the stellar-metallicity / planetary heavy-element correlation continues to hold with a larger number of planets.  This appears to fit well with expectations for the core-accretion theory of giant planet formation \cp{Dodson09}.  Determining whether this and perhaps other relations hold in the future will be an important area of transiting planet research in the future, as the number of detected planets continues to climb.

\bigskip
\noindent
\textbf{4.4 Proposed Solutions To The Radius Anomaly} \label{anom}
\bigskip

 As mentioned in \S 3.1, a significant fraction of transiting planets have large radii that cannot be explained by the effect of thermal irradiation
 from the parent star alone. Other mechanisms are required to explain these
 inflated planets.  Whether these mechanisms are peculiar, 
 operating only under specific conditions, for certain planets, or whether  basic
 physics is missing 
in the modeling of close-in giant planet structure are still open 
questions.  In either scenario the inflation mechanism(s) must be common.
 As noted, the latter alternative stems from the suspicion that all 
 observed transiting planets may contain a certain amount of heavy 
 material. The denser planetary structure resulting from the presence
of heavy material could counteract  the effect of  the ``missing" mechanism, yielding
an observed ``normal" radius \cp{Fortney06}.

 Many mechanisms have been proposed since the discovery
 of HD 209458b, but no mechanism has gained the consensus of the community as being clearly important. We describe below the main ideas
and comment on their status.

\begin{itemize}
\item{} {\em Atmospheric Circulation}

Based on numerical simulations of atmospheric circulation on hot Jupiters,
{\em Showman and Guillot} (2002) suggested a heating mechanism in the deep
interior driven by strong winds blowing on the planetary surface. The idea
is that stellar irradiation produces strong day-night temperature contrasts,
which drive fast winds. Their
simulations produced a downward kinetic energy flux, about 1\% of the absorbed
stellar incident flux, which dissipates in the deep layers. This mechanism
provides an extra source of energy which slows down the planet evolution
and can explain the large observed radii  ({\em Guillot and Showman}, 2002; 
{\em Chabrier et al.} 2004).

This attractive and original scenario still requires confirmation. The very existence of
the downward kinetic energy flux and its strength  strongly depend on 
the outcome of atmospheric circulation models. Although this field 
has almost half a century history for solar system planets, 
it is still in its infancy when applied to hot Jupiters where 
conditions are very different. Current models show important
divergences (see the chapter by Showman et al.)
and their robustness to describe the conversion process of
heat from the parent star into mechanical energy ({\em i.e} winds)
has even been questioned \cp{Goodman08}. The substantial downward
transport of kinetic energy reported by \ct{Showman02} has not been
found in somewhat similar numerical simulations performed by \ct{Burkert05}. Moreover, the physical process which converts 
kinetic energy into thermal energy in the {\em Showman and Guillot} scenario still needs to be identified.

\item{} {\em Enhanced Atmospheric Opacities}

\ct{Burrows07} suggest that abnormally
large radii of transiting planets can be explained by 
invoking enhanced opacities.  Enhanced  opacities could for instance be due to missing or underestimated opacities in the current generation of model atmospheres.  In practice, \ct{Burrows07} implement these higher opacities by simply using supersolar metallicities (up to 10 times
solar). The main effect of enhanced opacities is to retain internal
heat and to slow down the contraction. These authors conclude that
a combination of enhanced opacities and the presence of dense cores can 
provide a general explanation to the observed spread 
in radii of transiting planets. Interestingly, this work also finds
the same correlation as suggested by \ct{Guillot06}
between planet core mass and stellar metallicity.

If opacity calculations are indeed correct, then higher opacities could be due only to increased metallicity. This would, however, present several weaknesses. 
It may be difficult to maintain  a substantial fraction of
heavy elements in strongly irradiated atmospheres, which are radiative
down to deep levels.  Perhaps mixing via atmospheric dynamics suffices to keep the atmosphere well mixed \cp{Showman09,Spiegel09}. 
Moreover, as discussed in \ct{Guillot08}, \ct{Burrows07} do not take into account the subsequent increase 
of molecular weight due to the increase of the planet atmosphere 
metallicity. This effect counters the effect of enhanced opacities
and may even dominate in some cases, yielding the opposite effect on the radius
(see discussion in {\em Guillot}, 2008). Finally, for the most inflated
transiting planets yet detected, models with
enhanced opacities are unable to reproduce their radii and another
additional mechanism is required \cp{Liu08}.

\item{} {\em Tidal Effects}

Another alternative, first suggested by \ct{Bodenheimer01}, is that tidal forces may heat the planet. There has come to be a large body of work in this area.  If a planet has an eccentric orbit or
a non synchronous rotation, internal tidal dissipation within the planet produces
an energy source, which can slow down its contraction, or re-inflate the planet after a previous contraction phase \cp[see, e.g.][]{Gu03}. Following this idea, {\em Liu et al.}  (2008) suggest a combination of enhanced atmospheric opacities, the presence of heavy elements and heating due to orbital tidal dissipation, assuming non-zero
 eccentricities, to explain the most inflated planets. 
Almost all studies on tidal effects have assumed a tidal equilibrium state, yielding very short timescales for synchronization ($\sim 10^5-10^6$ yr) and circularization 
($\sim 10^8-10^9$ yr)  compared to the estimated age of 
known exoplanetary systems. This assumption implies that
for eccentricity to be maintained on several Gyr and contribute to the heating, it should be continually excited. In the case of HD 209458b, \ct{Bodenheimer01}  suggested the presence of an unseen planetary companion, which could force eccentricity.

Tidal heating due to a finite current eccentricity may explain some of the large
transit radii currently observed. It is, however, certainly not the mechanism
which could explain all inflated planets.  Constraints on the eccentricity
 of HD 209458b based on the timing of the secondary eclipse \cp{Deming05b}
 yield that a non-zero eccentricity is very unlikely. This explanation
seems also improbable for TrES-4 based on {\em Spitzer} observations by \ct{Knutson08c} who can rule out tidal heating at the level required by {\em Liu et al.} (2008) to explain this planet's bloated size. 

More recently, \ct{Jackson08a} and \ct{Levrard09} have revisited the tidal stability of exoplanets.  Essentially all examined transiting planets 
have not reached a tidal equilibrium state, implying that they will ultimately fall onto the
central star \cp{Levrard09}.  \ct{Jackson09}, confirming this view, find that it is the youngest parent stars that tend to harbor the closest-in hot Jupiters, implying a loss of close-in planets with time.  More importantly, these works stress that conventional circularization and synchronization timescales, which are widely used in the community, are in most cases not correct.  In the \ct{Levrard09} view, nearly circular orbits of planetary transiting systems currently observed may not be due to tidal dissipation.  Conversely, observations of non-zero eccentricity would be naturally explained without the need for gravitational interactions from undetected companions.
 
\ct{Ibgui09} and \ct{Miller09} have coupled a standard 2nd-order tidal evolution theory \cp[e.g.][]{Jackson09} to planet structural evolution models for close-in giant planets in single-planet systems to investigate under what circumstances tidal heating by recent eccentricity damping (which leads to an energy surge and radius inflation), together with tidal semimajor axis decay, could inflate these planets.  \ct{Miller09} find that this radius inflation can occur for some planets at Gyr ages, perhaps explaining some large radii, but this mechanism is unlikely to explain all inflated radii.

Tidal heating could also be produced by a large obliquity, {\em i.e} the angle between
the planetary spin axis and the orbital normal, as suggested by \ct{Winn05}. Since 
planet obliquity is expected to be rapidly damped by tidal dissipation, a possibility to maintain a non-zero obliquity is to be locked in a Cassini state, {\em i.e} a resonance between spin and 
orbital precession. However \ct{Levrard07} show that although the probability
of capture in a spin-orbit resonance is rather good around 0.5 AU, it decreases 
dramatically with semi-major axis. Also, \ct{Fabrycky07} rule out possible
drivers, such as the presence of a second planet, of a high obliquity Cassini state for HD 209458b. They conclude that very special configurations are required
for obliquity tides to be an important source of heating. 
On the whole, it thus seems difficult to invoke tidal dissipation as the mechanism which could explain all currently observed inflated exoplanets, although important quantitative effects have been identified.  Considerable work in this area continues.

\item{} {\em Double Diffusive Convection}

\ct{Chabrier07c} suggest that the onset of layered or
oscillatory convection, due to the presence of molecular weight gradients, 
can reduce  heat transport in planetary interiors and slows down the contraction,
providing an explanation for the large spread in radii of transiting planets.
The formation of layers is a characteristic of double-diffusive convection which
may occur in a medium where two substances diffuse at different rates. 
This is a well known process in oceans or salty lakes where the two substances
are heat and salt. Applied to the interior of planets, in the presence
of a compositional gradient, convection can break into convective layers 
separated by thin diffusive layers. The heat transport efficiency
is thus significantly reduced, because of the presence of multiple diffusive
layers, compared to the case of a fully homogeneous planet where convection
is assumed to be fully adiabatic. This process is similar to the so-called semiconvection
in stars \cp{Stevenson79b}. Based on a phenomenological approach,
and assuming a molecular weight gradient in the most inner part of
the planet, 
{\em Chabrier and Baraffe} (2007) show that a significant number of
diffusive layers strongly reduce the heat escape. They suggest that the
composition gradient is
inherited from the formation process, during accretion of planetesimals 
and gas, or due to core erosion.
The upshot is a significantly
inflated planet  compared to its homogeneous and adiabatic counterpart.

The idea that planetary interiors may not be completely homogeneous and 
convection not fully efficient was already suggested by D.~Stevenson for Jupiter \cp{Stevenson85} and for Uranus and Neptune \cp{Nbook}. (See \S3.)  Double diffusive convection is indeed a well known process
under Earth conditions, which have similarities with those found in the interior of giant planets (see {\em Chabrier and Baraffe}, 2007 for details). 
Whether this scenario can explain all inflated transiting planets is still debatable.  The key questions are whether the diffusive layers commonly form, and can survive on timescales of Gyr, characteristic of the age of the transiting exoplanets.  Development of 3D numerical simulations of this process under planetary conditions could 
provide clues on its long term existence in giant planet interiors. 

\end{itemize}
\begin{figure}
 \epsscale{0.8}
 \plotone{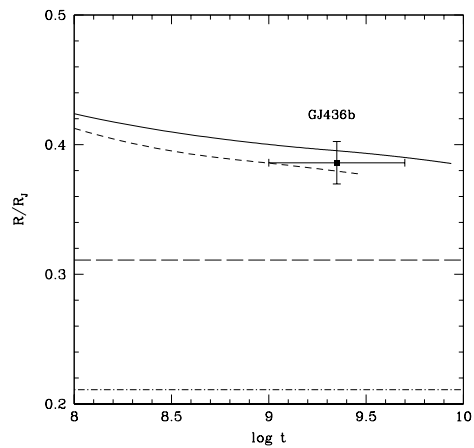}
  \caption{\small Evolution of a planet with the mass of GJ 436b (22.6 $\mearth$)
  and different heavy element contents and compositions.
  The solid line corresponds to a model with a water core of 21 $\mearth$ and the dashed line to a rocky core of 19.5 $\mearth$ (after the models of {\em Baraffe et al}, 2008). 
The long dashed line indicates the radius of a pure water planet and the dash-dotted line 
corresponds to a pure rocky planet (after the models of {\em Fortney et al.} , 2007)
}  \label{figgj436}
 \end{figure}
 
\bigskip
\noindent
\textbf{4.5 Hot Neptune Planets}
\bigskip

The dream of discovering extrasolar planets of a few Earth masses is now  becoming reality.
More than a dozen planets with masses in the Uranus-Neptune range ($\simle 20 \mearth$) has been detected by radial velocity surveys \cp{Udry07}.  
Because these light planets are very close to the detection threshold, their discovery suggest that they are rather common. Given their low mass and close orbit, the question of their origin deserves some attention. Formation models based on the core accretion scenario suggest that hot Neptunes are composed of a large heavy material
core (ice or rock) and have formed without accumulation of a substantial gaseous envelope \cp{Alibert06,Mordasini08}. Another suggestion is that hot Neptunes could have formed from more massive progenitors and have lost most of their gaseous envelope \cp{Baraffe05}. The latter idea arises from observational evidences that close-in exoplanets may undergo evaporation processes induced by the high energy flux of the parent star \cp{Vidal03}, as well as high mass-loss rates found in some early models \cp[e.g.][]{Lammer03}.

The interpretation of the \ct{Vidal03} observations of an extended neutral hydrogen atmosphere around \hd\ is an active area that is still controversial \cp{Ben07,Holmstrom08,Vidal08,Ben08}.  However, at the same time, the various groups computing models of evaporative mass loss at small orbital distances have been converging to mass loss rates that yield a total mass loss of only $\sim$~1\% for \hd\ over the lifetime of the system \cp{Yelle04,Tian05,Garcia07,Yelle08,Murray08}.  If this is correct, than these evaporation processes are too small to significantly affect the evolution of the \emph{currently observed} close-in planets.  Small evaporation rates seem also to be consistent
 with the observed mass function of exoplanets \cp{Hubbard07b}.  Improved statistics in the low planetary mass regime, confirmation of the  observations of {\em Vidal-Madjar et al.} (2003), and the extension of similar observations to other transiting planets will shed light on these issues.

Independently of these issues, the interior properties of two extrasolar hot Neptune planets were recently revealed by
the remarkable discovery of the first transiting Neptune-mass planets, GJ 436b \cp{Gillon07} and HAT-P-11b \cp{Bakos09}.  Both planets have a radius comparable to that of Neptune,
indicating heavy material enrichment greater than 85\%, which is the overall heavy element content of Uranus and Neptune ({\em Guillot}, 2005). Given current uncertainties on planetary interior structure models, only the bulk of heavy elements
can be inferred for each, about 20 $\mearth$ for a total mass of $\sim$ 22 $\mearth$ for GJ 436b, which is the more well-studied of the two (see Fig. \ref{figgj436}). Although small in terms of mass, the contribution of the H/He envelope to the total planetary radius is significant. This is  illustrated in Fig. \ref{figgj436}, where the radii of pure water and pure rocky planets of the same mass are also indicated.

Better determination of the chemical composition of heavy material and its distribution within the planet must await improved EOSs for H/He and heavy elements ({\em Baraffe et al.}, 2008). These discoveries however confirms the large heavy element content that can be expected in extrasolar planets and supports the general picture of planet formation drawn by the core accretion model.

\bigskip
\noindent
\textbf{4.6 Young Giant Planets} \label{youngj}
\bigskip

Giant planet thermal evolution models are being tested at Gyr ages for solar system planets and the transiting planets.  It is clear from giant planet formation theories (see the chapter by D'Angelo \& Lissauer) that these planets are hot, luminous, and have larger radii at young ages, and they contract and cool inexorably as they age.  However, since the planet formation process is not well understood \emph{in detail}, we understand very little about the initial conditions for the planets' subsequent cooling.  Since the Kelvin-Helmholtz time is very short at young ages (when the luminosity is high and radius is large) it is expected that giant planets forget their initial conditions quickly.  This idea was established with the initial Jupiter cooling models in the 1970s \cp{Graboske75,Bodenheimer76}.

Since our solar system's giant planets are thought to be 4.5 Gyr old, there is little worry about how thermal evolution models of these planets are affected by the unknown initial conditions.  The same may not be true for very young planets, however.  Since giant planets are considerably brighter at young ages, searches to directly image planets now focus on young stars.  At long last, these searches are now bearing fruit \cp{Chauvin05,Marois08,Kalas08,Lagrange09}.  It is at ages of a few million years where understanding the initial conditions and early evolution history is particularly important.  Traditional evolution models (\S \ref{evol}), which are applied to both giant planets and brown dwarfs, employ an arbitrary starting point.  The initial model is large in radius, luminosity, and usually fully adiabatic.  The exact choice of the starting model is often thought to be unimportant, if one is interested in following the evolution for ages greater than 1 Myr \cp{Burrows97,Chabrier00}.

Thermal evolution models, when coupled to a grid of model atmospheres, aim to predict the luminosity, radius, \teff, thermal emission spectrum, and reflected spectrum, as a function of time.  When a planetary candidate is imaged, often only the apparent magnitude in a few infrared bands are known.  If the age of the parent star can be estimated (itself a tricky task) then the observed infrared magnitudes can be compared with calculations of model planets for various masses, to estimate the planet's mass.  Recall that mass is not an observable quantity unless some dynamical information is also known. It is not known if these thermal evolution models are accurate at young ages--they are relatively untested, which has been stressed by \ct{Baraffe02} for brown dwarfs and \ct{Marley07} for planets.

\begin{figure}
 \epsscale{1.0}
 \plotone{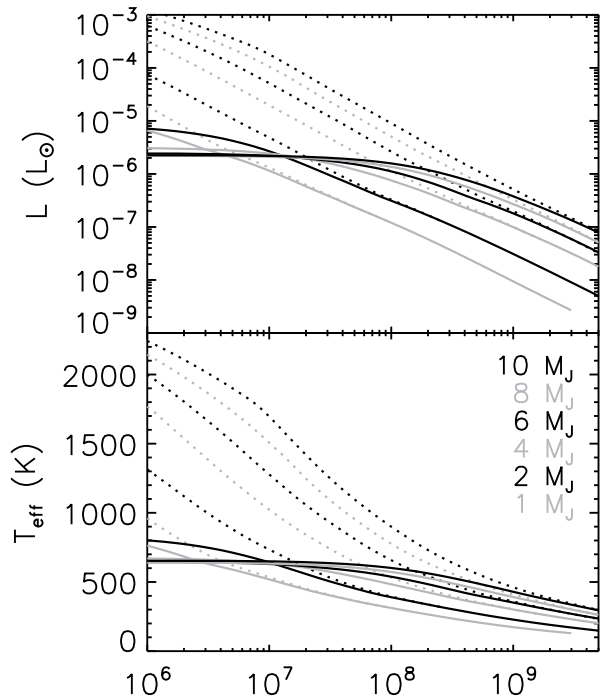}
  \caption{\small Thermal evolution of giant planets from 1 to 10 \mj, adapted from \ct{Marley07}.  The dotted curves are standard ``hot start'' models with an arbitrary initial condition, and the solid curves use as an initial condition the core accretion formation models of \ct{Hubickyj05}.
}  \label{young}
 \end{figure}

\ct{Marley07} examined the issue of the accuracy of the arbitrary initial conditions (termed a ``hot start'' by the authors) by using initial conditions for cooling that were not arbitrary, but rather were given by a leading core accretion planet formation model \cp{Hubickyj05}.  The core accretion calculation predicts the planetary structure at the end of formation, when the planet has reached its final mass.  The \ct{Marley07} cooling models used this initial model for time zero, and subsequent cooling was followed as in previously published models.  Figure \ref{young} shows the resulting evolution.  The cooling curves are dramatically different, yielding cooler (and smaller) planets.  The initial conditions are not quickly ``forgotten,'' meaning that the cooling curves do not overlap with the arbitrary start models for 10$^7$ to 10$^9$ years.  What this would mean, in principle, is that a mass derived from ``hot start'' evolutionary tracks would significantly underestimate the true mass of a planet formed by core accretion.

Certainly one must remember that a host of assumptions go into the formation model, so it is unlikely that these new cooling models are quantitatively correct.  However, they highlight that much additional work is needed to understand the \emph{energetics} of the planet formation process.  The \ct{Hubickyj05} models yield relatively cold initial models because of an assumption that accreting gas is shocked and readily radiates away this energy during formation.  This energy loss directly leads to a low luminosity starting point for subsequent evolution.  Significant additional work on multi-dimensional accretion must be done, as well as on radiative transfer during the accretion phase, before we can confidently model the early evolution.  Thankfully, it appears that detections of young planets are now beginning to progress quickly, which will help to constrain these models.
\begin{figure}[htp]
 \epsscale{1.0}
 \plotone{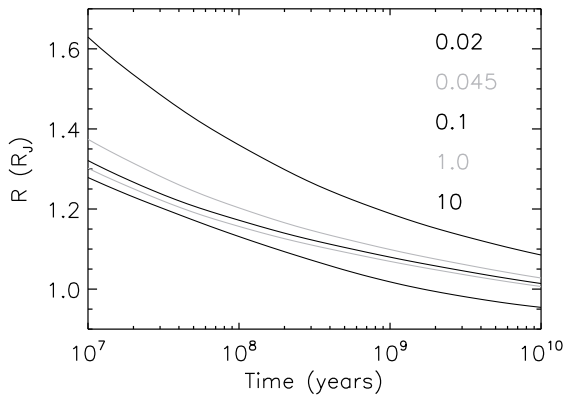}
  \caption{\small Contraction of a 1 \mj\ planet, with 25 \me\ of heavy elements (in a core), over time.  The planets are placed at 0.02, 0.045, 0.1, 1.0, and 10 AU from a constant luminosity Sun, to show the effect of stellar irradiation on thermal evolution.  Adapted from \ct{Fortney07a}.
}  \label{dist}
 \end{figure}

\bigskip
\section{FUTURE PROSPECTS}
\bigskip

The future of understanding the structure, composition, and evolution of giant planets is quite promising.  Most immediately, and with the biggest impact, will be the detection of more Neptune-class transiting planets, in addition to GJ 436b and HAT-P-11b.  Of particular interest will be the mass-radius relation of planets around $\sim$10 \me, since this is estimated to be a boundary between planets that have H/He envelopes and those that do not.  A radius that is larger than that calculated for a pure water planet is unambiguous evidence of a H/He envelope.  Certainly the diversity of radii for the transiting gas giants has been surprising, and understanding Neptune-class radii as a function of mass and orbital distance will be fascinating.

Several ground-based surveys for transiting planets are now scouring the northern hemisphere for Jupiter-class planets, with some additional attention being paid to the southern hemisphere.  The required photometric precision to detect a Jupiter-type transit (a 1\% dip in stellar flux) is not difficult to achieve.  Furthermore, the detection of HAT-P-11b opens up the possibility that these same ground-based surveys will also be able to detect significant numbers of Neptune-type planets as well.  In addition, the implementation of orthogonal transfer array CCDs may allow for ground-based photometric precision that approaches that of space-based platforms \cp[e.g.][]{Johnson09}, which would significantly help the cause of detecting smaller planets from the ground.

The \emph{CoRoT} mission, which has already announced five planets, and will continue to search for a few more years, will be important in adding to our sample size in two ways.  The first is of course smaller planets.  The second is increasing the sample size of transiting planets at wider orbital separations.  \emph{CoRoT} is surveying particular areas of the sky for 120 days, meaning that 40-day orbits of giant planets can be easily confirmed via 3 detected transits.  Around a solar-type star, a 40-day orbit yields a semi-major axis of 0.23 AU, which is a factor 33 reduction in flux from that intercepted at 0.04 AU.  Understanding radii at the largest possible range of incident fluxes may allow us to understand the reason(s) for the large radii of the currently detected planets.

The \emph{Kepler} mission, which began taking science data in May 2009, is an even more ambitious telescope to detect smaller transiting planets in longer period orbits.  Although its main goal is to ascertain the frequency of Earth-radius planets in Earth-like orbits around Sun-like stars, it will also do the same for larger planets.  \emph{Kepler} will be able to detect multiple transits of planets out to 1 AU, a factor of over 600 reduction in flux from 0.04 AU.  In addition, even longer period transiting giant planets may be detected, if suitable followup is done.  Since some mechanisms proposed to explain the large radii of the close-in planets should be significantly muted at larger orbital separations, finding planets farther from their parent stars will likely be the \emph{most important} step in understanding what leads to these large radii.  In Figure \ref{dist} we show a specific prediction for the contraction of a 1 \mj\ planet over a factor of 250,000 in incident stellar flux.  As detailed in \ct{Fortney07a}, the effects of stellar flux are muted beyond the current group of close-in transiting planets, and planets out to $\sim$1 AU should have radii quite similar to those at $\sim$0.1 AU, in the absence of missing physics for these more distant planets.

The orbital dynamics of particular exoplanets in some systems may give us direct constraints on a planet's interior state.  The apsidal precession rate of a planetary orbit is directly proportional to the tidal Love number, $k_2$.  This number parameterizes the internal density distribution of a planet.  Authors have recently pointed out instances in which $k_2$ could be measured or well-constrained.  These include precession due to a tidal-induced gravitational quadrupole on the planet by its parent star, for very close-in planets \cp{Ragozzine09}, which could be measured as a change in transit shape over time.  Another affects transiting planets in multi-planet systems, which now only includes HAT-P-13b \cp{Bakos09b}.  For this system a refined measurement of current planetary eccentricity can constrain $k_2$ as well as $Q$, its tidal dissipation quality factor \cp{Batygin09}.

Detections of transiting giant planets at younger ages would be very important as these planets would inform our understanding of contraction with time in the face of intense stellar irradiation.  This would shed light on the initial conditions for evolution, post-formation, as well as allow us to better understand the nature of the physical process that is causing large radii at gigayear ages.  Towards this goal, several transit surveys of open clusters have been performed, but they have not netted any planets to date.

The focus of comparing models to observations is already shifting from \emph{specific planets} to \emph{samples of planets}, and will soon shift to a statistically significant number of planets, with the additional detections from the ground and from space.  A good reference for the kind of work, just starting to be done, is \ct{Fressin07}, who analyzed in detail the OGLE transiting planet survey.  They simulate the OGLE survey: given the properties of the thousands of stars that were monitored, the known planet frequency as a function of stellar mass and of orbital distance, and implement giant planet contraction models, to derive constraints on, for instance, possible separate populations of planets. 

The planets directly imaged by \ct{Marois08} and \ct{Kalas08} have fully opened the door to direct imaging, which began yielding planetary candidates a few years ago \cp[e.g.][]{Chauvin05}.  The characterization of these planets will present different challenges compared to the transiting planets.  While transiting planets yield accurate masses and radii, the atmospheric characterization is challenging.  For directly imaged planets, masses and radii likely cannot be measured, but spectra should be more easily obtained.  Spectra can yield the planet's \teff, which can be directly compared to thermal evolution models.  However, current techniques are limited to planet-to-star flux ratios of $\sim10^{-5}$.  New instruments coming online in the very near future, such as the Gemini Planet Imager (GPI) on Gemini South and Spectro-Polarimetric High-contrast Exoplanet REsearch (SPHERE) at the VLT will allow for contrasts of $\sim10^{-6}-10^{-7}$.  This will allow for the direct imaging and characterization of giant planets at a variety of masses, and also a variety of \emph{ages}, so the first $\sim1-100$ Myr of giant planet evolution should be relatively well understood via observations.

We are now over a decade into the new era of studying giant planets as a class of astronomical objects.  It is the expansion of this class beyond Jupiter, Saturn, Uranus, and Neptune, that will enable us to better understand the formation and evolution of these planets.  Much like our understanding of the stars is greatly enhanced by studying more than just the Sun, so will our understanding of giant planets grow.

Over the past several years we have seen strange and startling transiting planets with hugely inflated radii, and those with small radii that must include vast interior stores of heavy elements.  They have expanded our imaginations regarding the possible structure of giant planets.  We are now beginning to see these original oddballs within the continuum of giants planets that extend far beyond what we see in our solar system.  Astronomers will do doubt continue to creatively find ways to detect and characterize many more planets in the future.

\bigskip
\textbf{ Acknowledgments.} JJF was partially supported by the National Science Foundation and the NASA Outer Planets Research Program.

\bigskip

\bigskip
\parskip=0pt
{\small
\baselineskip=11pt


\end{document}